\documentclass[referee]{raa}            
\usepackage{graphicx,times}             
\usepackage{natbib}
\usepackage{amssymb,amsmath}

\newcommand\period{$days\cdot{cycle}^{-1}$}

\begin{document}

\title{ Multi-color light curves and orbital period research of eclipsing binary V1073 Cyg }

\volnopage{Vol.0 (200x) No.0, 000--000}      
   \setcounter{page}{1}          

\author{
        Xiao-man Tian
        \inst{1, 2, 3}
       \and Li-ying Zhu
       \inst{1, 2, 3}
       \and Sheng-bang Qian
       \inst{1, 2, 3}
        \and Lin-jia Li
       \inst{1, 2, 3}
       \and Lin-qiao Jiang
       \inst{4}
        }

\institute{
  Yunnan Observatories,   Chinese Academy of Sciences (CAS),   P. O. Box 110,  650216 Kunming,   China; {\it joie@ynao.ac.cn}\\
\and
  Key Laboratory of the Structure and Evolution of Celestial Objects,  Chinese Academy of Sciences, P. O. Box 110, 650216 Kunming,  China\\
\and
 University of the Chinese Academy of Science,   Yuquan Road 19,  Sijingshang Block,  100049 Beijing, China\\
\and
School of Physics and Electronic Engineering, Sichuan University of Science $\&$ Engineering, Zigong 643000, China\\
}
\date{Received~~2017 month day; accepted~~2017 month day}

\abstract{
New Multi-color $B$ $V$ $R_c$ $I_c$ photometric observation are presented for W UMa type eclipsing binary V1073 Cyg. The multi-color light curves analysis with the Wilson-Devinney(W-D) procedure acquired the absolute parameters of this system, showing that V1073Cyg is a shallow contact binary system with fill-out factor $f=0.124(\pm0.011)$. We collected all available times of light minima spanning 119 years including CCD data to construct the O-C curve and made detailed O-C analysis. The O-C diagram shows that the period change is complex. There exist a long-term continuous decrease and a cyclic variation. The period is decreasing at a rate of $\dot P=-1.04(\pm0.18)\times 10^{-10}$\period, and with the period decrease, V1073 Cyg will evolve to deep contact stage.
The cyclic variation with a period of $P_3=82.7(\pm3.6) years$ and an amplitude of $A=0.028(\pm0.002) day$ may be explained by the magnetic activity of one or both components or the light travel time effect (LTTE) caused by a distant third companion with $M_3({i'}=90^{\circ})=0.511M_\odot$.
}

  \authorrunning{Tian,et,al. }            
   \titlerunning{Research of eclipsing binary V1073 Cyg }  
\maketitle

\section{Introduction}
V1073 Cyg (BD $+33^{\circ}$ 4252, HD 204038, HIP 105739, BV 342) is a W UMa type eclipsing binary. \citet{strohmeier1960} first recognized the variability of V1073 Cyg, then the photographic light curve was published in 1962 \citep{strohmeier1962}. Thus far, more photometric and spectrographic researches have been carried out for V1073 Cyg (such as \citet{sezer1993, morris2000, yang2000,ekmekci2012} and so on), and absolute parameters have been derived. Here are the brief introduction. Radial velocity curve was first reported by \citet{fitzgerald1964}, the mass ratio $q=0.34$ was obtained. The latest mass ratio $q=0.303(17)$ was presented by \citet{pribulla2006} based on their radial-velocity curve. And the fill-out factor has been reported many times, such as $f=0.007$ and $f=0.008$\citep{kondo1966,leung1978,ahn1992,sezer1993,sezer1994}, $f=0.19$ \citep{jafari2006}, and $f=0.20$ \citep{sezer1996}. The original classification of the spectral type of the primary star was A3 Vm\citep{fitzgerald1964}, some other spectral types classification also were reported as F0n III-IV\citep{abt1969}, F0n V\citep{morgan1943} and F0V \citep{pribulla2006}. Investigations about V1073 Cyg have confirmed that this system is a short-period contact eclipsing system with a period of $P=0.7858506 day$ \citep{pribulla2006}.

The orbital period analysis has been taken out by many authors. \citet{aslanherczeg1984} found that the period decreased about 0.4 seconds in 1976 using 25 photoelectric and some photograph epochs of minima since 1962, then \citet{wolf1992} made a new O-C analysis with adding 16 photoelectric times of light minima and detected a constant decrease.
One year later, \citet{sezer1993} made a analysis using 29 photoelectric data and reported that the period was decreasing by 3.12($\pm$0.17) seconds per century (i.e. $\dot P=-7.8\times{10}^{-10}$ \period). 17 years ago, \citet{morris2000} pointed out that the period had decreased $0.795\pm 0.040s$ around JD 2,445,000 in 1982, while \citet{yang2000} found that the period was decreasing at a rate of $\dot P=-8.8\times{10}^{-10}$ \period from 1962 to 1998 year, using 111 data including 62 pe data in their analysis. By now, more high-precision eclipse times have been obtained, which are very useful for more precise period investigation.

According to the spectroscopic observation made by \citet{fitzgerald1964}, the spectrum of the primary component shows that the primary component of V1073 Cyg is a Am star. V1073 Cyg has been listed in the catalog of Ap, HgMn and Am stars \citep{renson2009} as a doubtful Am star. Am stars almost are A and F type stars with remarkable peculiar characteristic of element abundances, such as considerably weaker Ca II, K line \citep{titus1940, roman1948}, under-abundance of calcium and scandium, over-abundances of iron-group elements, and extreme enhancements of rare-earth elements \citep{conti1970} compared to the same spectral type stars. Am stars rotate more slow than normal A and F type stars, and the rotational velocities are always less than about $100 km\cdot s^{-1}$ \citep{abt1973};
Am stars show high binaries-proportion, which is more than $90\%$ \citep{abt1961, abt1965, hubrig2010}. As the only one W UMa type binary of the 73 eclipsing Am binaries \citep{renson2009, smalley2014}, V1073 Cyg is very special because the rotation velocity of the primary component is about 160 $km\cdot s^{-1}$ \citep{pribulla2006}, which is much higher than the limit rotation velocity $100 km\cdot s^{-1}$ given by \citet{abt1973}, that makes it became a challenge to the cut-off of rotation velocity. So the investigations both photometric and spectroscopic of this target are very important.

In this study, we carried out a new photometric solution for V1073 Cyg based on our new multi-color light curves and derived new absolute parameters. We also collected the times of minima of V1073 Cyg spanning 119 years and made a new exhaustive orbital period analysis.

\section{OBSERVATION AND DATA REDUCTION}
\begin{figure}
  \begin{center}
   \includegraphics[width=12cm]{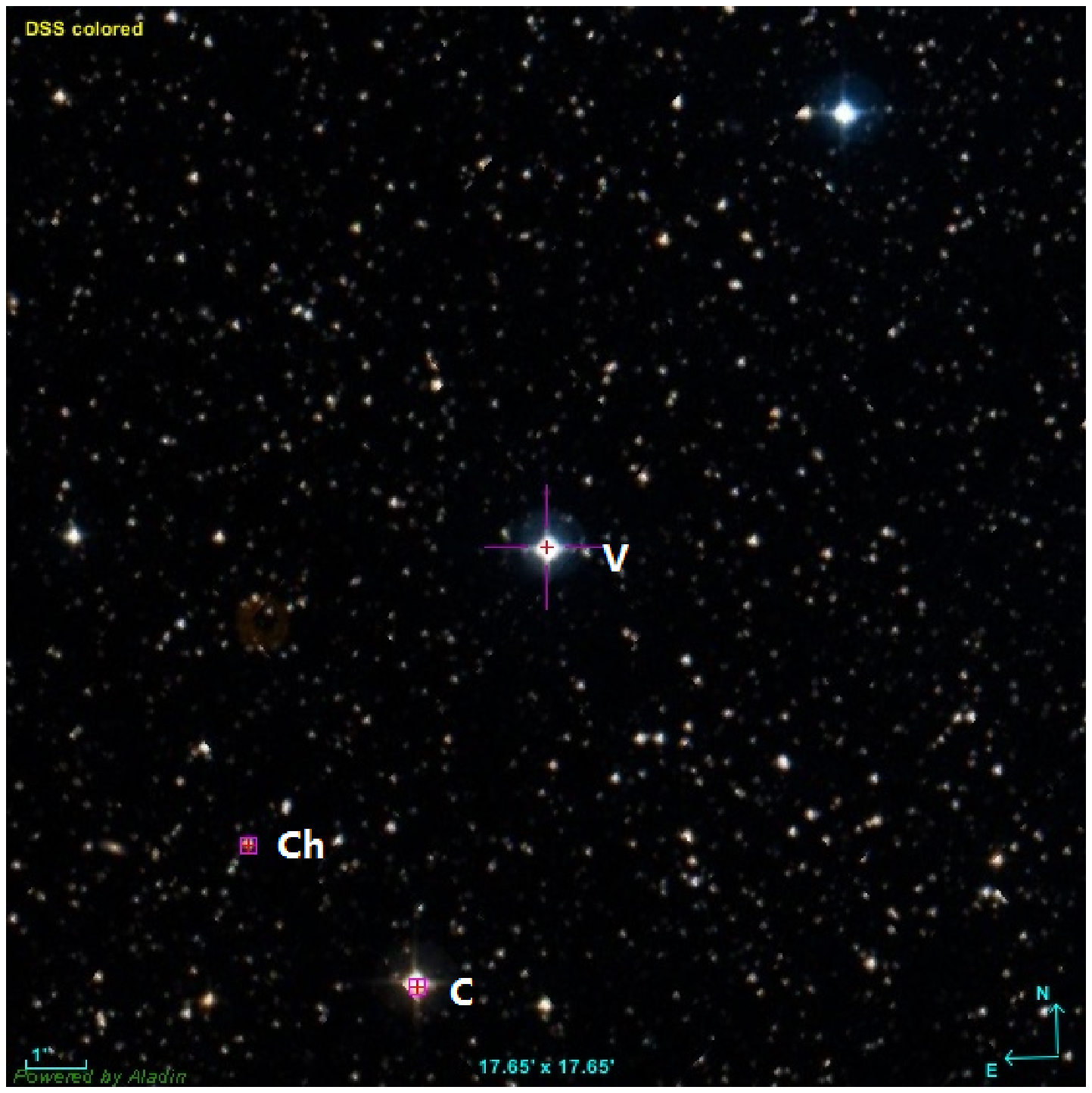}
  \end{center}
  \caption{Finding chart. 'V','C' and 'Ch' represent 'Variable star','Comparison star' and 'Check star' respectively.}
\label{fig1}
\end{figure}

Four filters light curve observations of V1073 Cyg were carried out on 2015 October 26, 27, 28, 29 using the Andor DZ936 2K CCD (size: $2048\times 2048$ pixel) photometric system attached to the 85 $cm$ telescope (labeled as $'D80 cm'$) at the Xinglong Station of National Astronomical Observatories of China (NAOC). And observations of minima light times were carried out on 2015 June 17 using $'D80 cm'$ and on 2015 September 28 using the Andor DW436 2K CCD camera mounted in the 60 $cm$ reflecting telescope (labeled as $'D60 cm'$) at the Yunnan Observatories in China (YNO).
The images were processed with the PHOT package of IRAF in a standard mode. Then, differential magnitudes were determined, by choosing BD+32 4152 as the comparison star and TYC 2707-276-1 as the check star. The relevant information are listed in Table \ref{table1} and the finding chart is shown in Figure \ref{fig1}.

\begin{table}
\caption{The information of V1073 Cyg, comparison and check stars}
\centering
\normalsize
\label{table1}
\begin{tabular}{lcccc}
\hline
      &   Variable(V)  &  Comparison(C)     &  Check(Ch)     \\
      & V1073 Cyg & BD+32 4152 &TYC 2707-276-1 \\
\hline
 ${\alpha}_{2000}$&  ${21}^{h}{25}^{m}{00}^{s}. 35766$   & ${21}^{h}{25}^{m}{10}^{s}. 6007 $ &${21}^{h}{23}^{m}{10}^{s}. 72$  \\
 ${\delta}_{2000}$ &   ${+33}^{\circ}{41}^{'}{14}^{''}. 9435$ & ${+33}^{\circ}{34}^{'}{09}^{''}. 306$ &${+33}^{\circ}{23}^{'}{29}^{''}. 9 $ \\
 V(mag) & 8. 38  & 8. 77 & 12.18  \\
\hline
\end{tabular}
\end{table}

\section{LIGHT CURVES AND ANALYSES}
\begin{figure}
  \begin{center}
   \includegraphics[width=12cm]{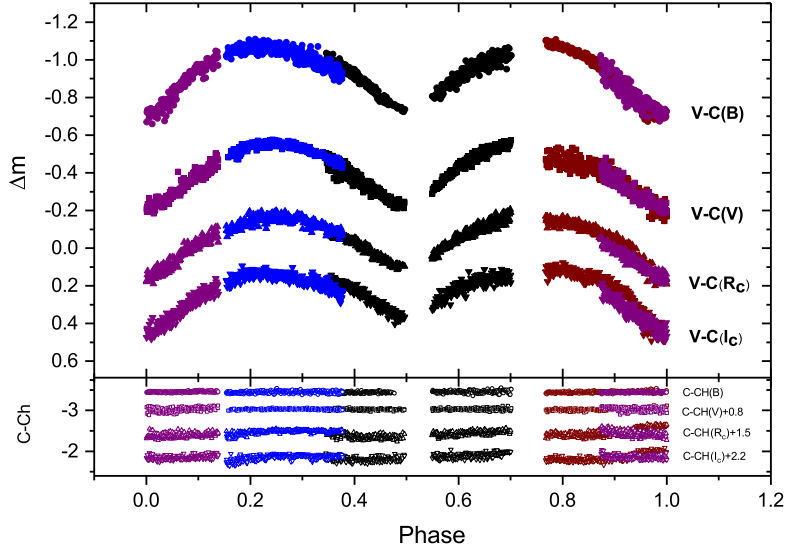}
  \end{center}
  \caption{The phased light curves of V1073 Cyg.
  Circle, square, triangle and down triangle represent B V $R_c$ $I_c$ bands observation data and the open symbols show the corresponding magnitude differences between comparison star and check star.}
\label{fig2}
\end{figure}

New multi-color curves of V1073 Cyg were obtained in this study. The photometric phases were calculated with the new linear ephemeris:
\begin{eqnarray}
Min \emph{I}& = &2457191.2159(.0003)+{0.}^{d}7858506\ast E
\end{eqnarray}
in which, 2457191.2159 is the new time of light minima obtained by this study, and $P=0.7858506 day$ was given by \citet{pribulla2006}, $E$ is the cycle number. The phased multi-color light curves are plotted in Figure \ref{fig2} with different color marking the observations of different day, and corresponding magnitude differences between comparison star and check star are shown in the bottom panel.

\begin{figure}
 \begin{center}
  \includegraphics[width=12cm]{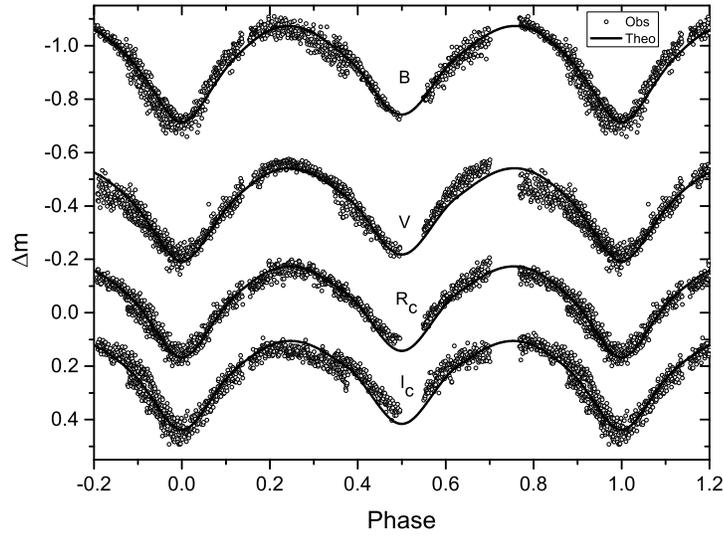}
  \end{center}
  \caption{The observed (circles) and theoretical light curves (solid lines) of V1073 Cyg.}
  \label{fig3}
\end{figure}

\begin{figure}
  \begin{center}
   \includegraphics[width=12cm]{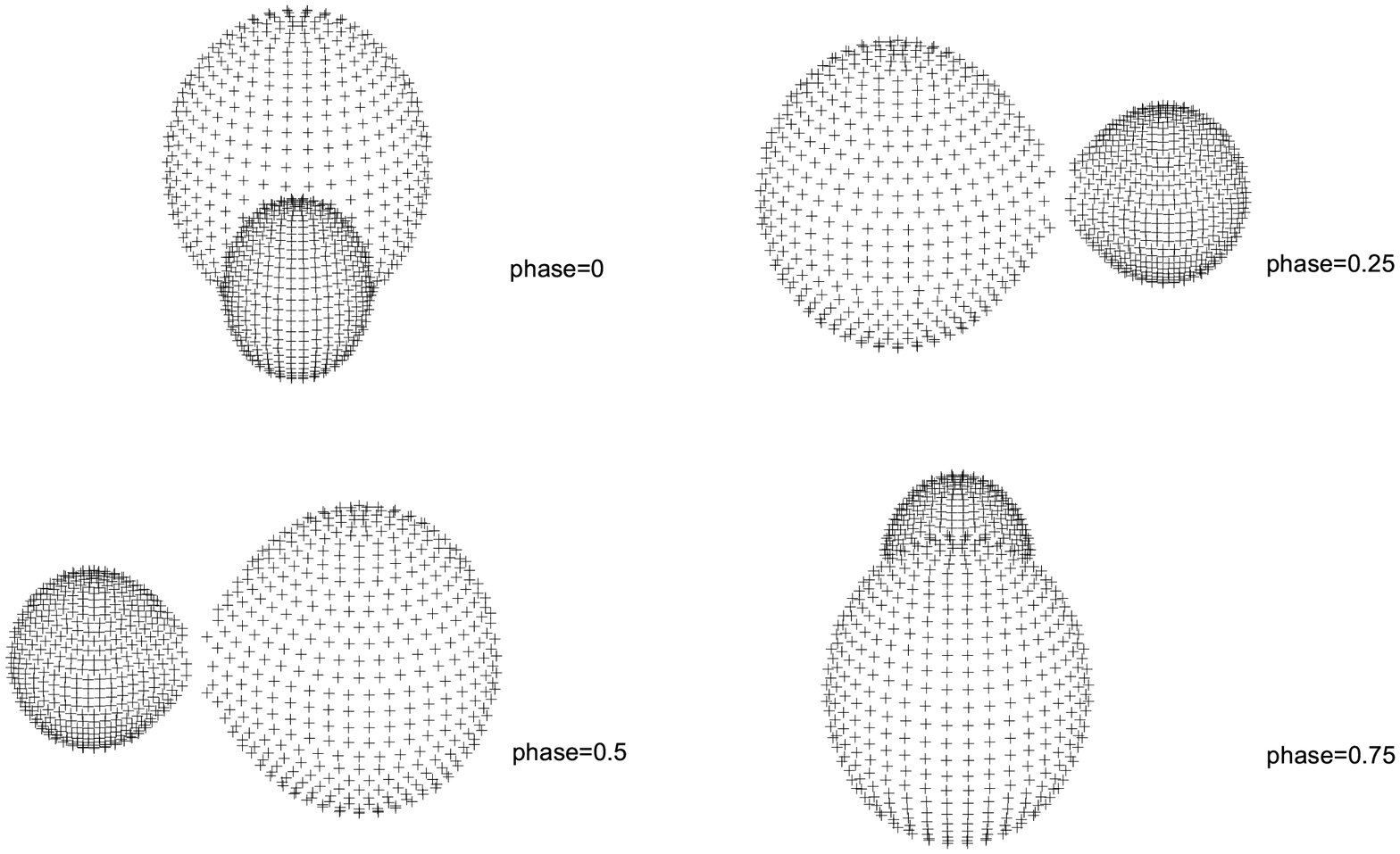}
  \end{center}
  \caption{ Geometrical structure of V1073 Cyg at phases 0.0,  0.25,  0.5 and 0.75 respectively.}
  \label{fig4}
\end{figure}

Absolute parameters of this system were derived from the light curves analysis with Mode 3 in the Wilson-Devinney(W-D) program \citep{wildev1971,wilson1979, wilson1990,van2007, wilson2008, wilson2010,wilson2012}. In the solution process, we adopted q=0.303 and F0 V as the spectral type of primary star \citep{pribulla2006}. Then refer to the Table 15.7 (Calibration of MK spectral types) in the book named Allen's Astrophysical Quantities by \cite{cox1999}, the temperature of primary star (star 1 in Mode 3) should be $T_1=7300K$, the convergent photometric solutions are derived and listed in Table \ref{table2}. We tried to set the third light $l_3$ as a free parameter, but failed to obtain the convergent solution.
Based on the Kepler's third law and $(M_1+M_2){\sin{i}^3}=1.896(25)$ \citep{pribulla2006}, the absolute parameters ($M_1$, $M_2$, $R_1$, $R_2$, $a$) can be derived, which are listed in Table \ref{table2}. The theoretical light curves are plotted in Figure \ref{fig3}, while the geometrical structure is shown in Figure \ref{fig4}.

\begin{table}{}
\centering
\normalsize
\caption{Photometric Solutions of V1073 Cyg}
\label{table2}
\begin{tabular}{lcc}
\hline
Parameter                    & Value                           \\
\hline
 Mode                        & 3                              \\
 q$(M_2/M_1)$                &0.303(fixed)                   \\
 $M_1(M_\odot)$              &1.810 ($\pm0.004$)            \\
 $M_2(M_\odot)$              &0.549 ($\pm0.001$)            \\
 $R_1(R_\odot)$              &2.545($\pm0.008$)                  \\
 $R_2(R_\odot)$              &1.481($\pm0.009$)                 \\
 $a(R_\odot)$                &5.172($\pm0.030$)                    \\
 $i(^\circ)$                 & 68.4($\pm0.1$)                  \\
 $T_1$(K)                    & 7300                             \\
 $T_2$(K)                    & 6609($\pm18$)                      \\
 $Omega_1$                   & 2.450($\pm$ 0.002)                   \\
$L_1/L_{total}(B)$           &0.832($\pm0.002$)                      \\
$L_1/L_{total}(V)$           &0.829($\pm0.002$)                       \\
$L_1/L_{total}(R_c)$         &0.804($\pm0.002$)                     \\
$L_1/L_{total}(I_c)$         &0.794 ($\pm0.003$)                     \\
$f(\emph{fill-out})$         &0.124 ($\pm$0.011)                        \\
$r_1(pole)$                  &0.460 ($\pm0.0005$)                  \\
$r_1(side)$                  &0.496 ($\pm0.0006$)                  \\
$r_1(back)$                  &0.522 ($\pm0.0008$)                   \\
$r_2(pole)$                  &0.267 ($\pm0.0005$)                    \\
$r_2(side)$                  &0.279 ($\pm0.0006$)                    \\
$r_2(back)$                  &0.315 ($\pm0.0010$)                     \\
$\sum(O-C)^2$                &0.035                                       \\
\hline
\end{tabular}
\end{table}

\section{O-C DIAGRAM ANALYSES}

\begin{figure}
  \centering
  \includegraphics[width=12cm]{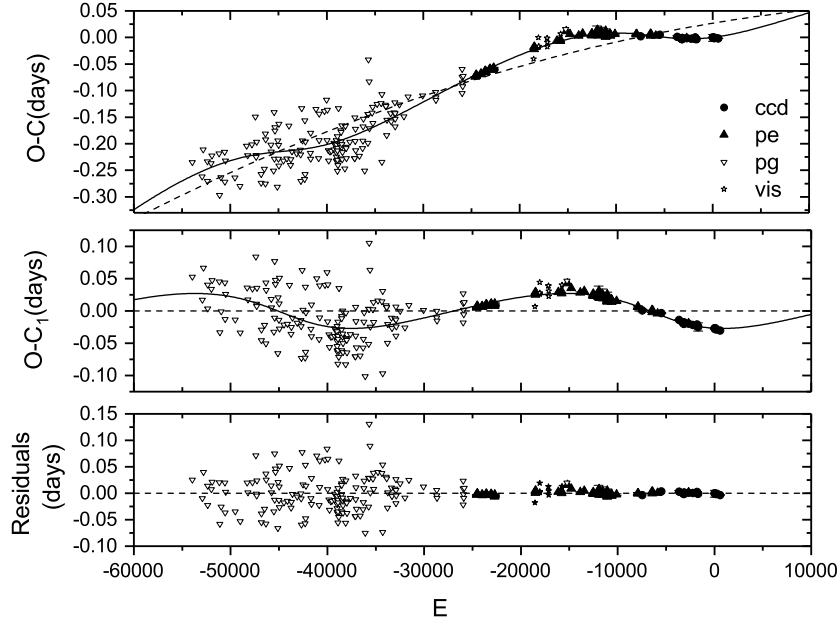}
  \caption{O-C diagram of V1073 Cyg. Spots, triangle, open down triangle and pentagram mark the $CCD$, $pe$, $ph$ and $vis$ data respectively. The solid line showing in the upper panel is the combination of the continuous decrease and the cyclic variation, while the sine-like curve showing in the middle panel refers to the cyclic variation $(O-C)_1$ after the subtraction of the continuous decrease, and the residuals are shown in the bottom panel.}
  \label{fig5}
\end{figure}

We collected 257 times of light minima from 1899 to 2017 and observed three times in 2015. All these 260 times of light minima were listed in Table \ref{table3}. The time HJD 2447408.7690 observed by spectrographic method was abandoned because of the obvious deviation from the O-C diagram, and all the others listed in the Table \ref{table3} were used for the O-C analysis. The O-C values were determined by the observed (O) mid-eclipse times minus the calculated (C) times with Ephemeris(1). The corresponding O-C diagram is shown in the upper panel of Figure \ref{fig5} with spot, triangle, pentagram and down triangle representing the $CCD$, $pe$, $vis$ and $ph$ data respectively.

The period decrease had been well discussed \citep{aslanherczeg1984,wolf1992,sezer1993,morris2000,yang2000}. We tried to fit the O-C curve with downward parabola curve indicating period decrease, which is shown with dished line in the upper panel of Figure \ref{fig5}. The residuals $(O-C)_1$ from the parabolic fit are displayed in the middle panel, which also show a cyclic oscillation indicating the existence of a cyclic variation. So we combined the downward parabola curve with a cyclic variation and tried to make a good fit to the general trend of the O-C curve. The combination curve is plot in the upper panel of Figure \ref{fig5} with solid line.
 Taking the case (eccentric orbital) reported by \citet{qian2015} into account, following equation (2) was used to describe the O-C diagram:
\begin{eqnarray}
\centering
 O-C &=& \triangle T_{0}+ \triangle P_{0} \times E+1/2\beta{E}^{2}+\tau
\end{eqnarray}
$\triangle T_{0}$ is the  the corresponding correction values of the epoch and $\triangle P_{0}$ is the period in Equation(1) , $\beta$ is the rate of the linear change of the period (\period). $\tau$ is the periodic variation caused by the light-travel time effect (LTTE) \citep{irwin1952a}, shown as $(O-C)_{1}$ in the middle panel of Figure \ref{fig5},
\begin{eqnarray}
\centering
 \tau &=& A[(1-e^2)\displaystyle\frac{\sin(\nu+\omega)}{1+e\cos(\nu)}+e\sin(\omega)]\\
        &=& A[\sqrt{(1-e^2)} \sin E^*\cos\omega+\cos E^*\sin\omega]
\end{eqnarray}
in which, $A={a_{12}}^{'} \sin{i^{'}}/c$ ($day$) is the projected semi-major axis and $c$ is the velocity of light, $e$, $\nu$, $\omega$ are the eccentricity, true anomaly and longitude of the periastron passage respectively , $E^*$ is the eccentric anomaly.
The Kepler equation leads to the relationship formula of the mean anomaly M and $E^*$:
\begin{eqnarray}
\centering
 M &=&E^*-e\sin E^*=\displaystyle\frac{2\pi}{P_A}(t-T)
\end{eqnarray}
$P_A$ is the anomalistic period, $t$ is the time of light minimum, and $T$ is the time of periastron passage respectively.

In our O-C analysis, we adopt different weight for different-precision data, i.e. a weight of 5 for CCD (ccd) and photoelectric (pe) data and a low weight of 1 for visual (vis) and photographic (pg) data, as same as that of \citet{yang2000}. The combination of the change of continuum decrease and cyclic variation gives a good fit to the O-C curve.
All the fitting curves have been shown in Figure \ref{fig5}, the solid line showing in the upper panel refers to the combination of the continuous decrease with a rate of $\dot P=-1.04(\pm0.18)\times 10^{-10}$\period and the cyclic variation with a period of $P_3=82.7(\pm3.6) years$ and an amplitude of $A=0.028(\pm0.002) day$.
While the curve showing in the middle panel is the cyclic variation $(O-C)_1$ after subtracting the continuous decrease, and the residuals are shown in the bottom panel. And all results obtained from the above period analysis are listed in Table \ref{table4}.

\begin{table*}
\caption{The times of light minima of V1073 Cyg}\label{table3}
\tiny
\centering
\begin{tabular}{llccllc|llccllc}
\hline
HJD &      & &   &      &       &    &   HJD &      & &   &      &       &      \\	
(2, 400, 000+) &Error& Min. & Meth. & E & O-C(day) &Ref.&(2, 400, 000+) &Error& Min. & Meth. & E & O-C(day) &Ref. \\
\hline
14801.8070 &  &s	&  pg	  &  -53940.5   &  -0.23464   &  1    &    23641.8050   &  &s	&  pg	  &  -42691.5   & -0.27004  &   1  \\
15601.7760 &  &s	&  pg	  &  -52922.5   &  -0.26155   &  1   &     23725.6350   &  &p	&  pg	  &  -42585     & -0.13313  &   1  \\
15694.5570 &  &s	&  pg	  &  -52804.5   &  -0.21092   &  1   &     23773.5110   &   &p	&  pg	  &  -42524     & -0.19402  &   1  \\
16029.6630 &  &p	&  pg	  &  -52378     &  -0.27021   &  1  &      23953.7910   &  &s	&  pg	  &  -42294.5   & -0.26673  &   1  \\
16033.6290 &  &p	&  pg	  &  -52373     &  -0.23346   &  1  &      23981.7990   &   &p	&  pg	  &  -42259     & -0.15643  &   1  \\
16330.6770 &  &p	&  pg	  &  -51995     &  -0.23698   &  1  &      24003.7020   &  &p	&  pg	  &  -42231     & -0.25724  &   1  \\
16413.5990 &  &s	&  pg	  &  -51889.5   &  -0.22222   &  1   &     24016.7300   &   &s	&  pg	  &  -42214.5   & -0.19578  &   1  \\
16990.8100 &  &p	&  pg	  &  -51155     &  -0.21849   &  1  &      24422.6000   &   &p	&  pg	  &  -41698     & -0.21761  &   1  \\
17021.7730&  &s	&  pg	  &  -51115.5   &  -0.29659   &  1  &      24459.5240   &   &p	&  pg	  &  -41651     & -0.22859  &   1  \\
17145.5780 &  &p	&  pg	  &  -50958     &  -0.26306   &  1  &      24459.5730   &   &p	&  pg	  &  -41651     & -0.17959  &   1  \\
17432.8540 &  &s	&  pg	  &  -50592.5   &  -0.21545   &  1   &     24889.4780   &  &p	&  pg	  &  -41104     & -0.13487  &   1  \\
17458.7310 &  &s	&  pg	  &  -50559.5   &  -0.27152   &  1   &     25102.7440   &  &s	&  pg	  &  -40832.5   & -0.22731  &   1  \\
17497.6700 &  &p	&  pg	  &  -50510     &  -0.23213   &  1  &      25132.6900   &  &s	&  pg	  &  -40794.5   & -0.14363  &   1  \\
18283.4890 &  &p	&  pg	  &  -49510     &  -0.26373   &  1   &     25139.7130   &  &s	&  pg	  &  -40785.5   & -0.19329  &   1  \\
18683.4710&  &p &  pg	  &  -49001     &  -0.27968   &  1   &     25573.5570   &  &s	&  pg	  &  -40233.5   & -0.13882  &   1  \\
19238.7500 &  &s	&  pg	  &  -48294.5   &  -0.20413   &  1   &     25748.8230  &  &s	&  pg	  &  -40010.5   & -0.11750  &   1  \\
19308.6730 &  &s	&  pg	  &  -48205.5   &  -0.22183   &  1   &     25779.7880   &  &p	&  pg	  &  -39971     & -0.19360  &   1  \\
19668.6130 &  &s	&  pg	  &  -47747.5   &  -0.20141   &  1   &     25803.8140  &  &s	&  pg	  &  -39940.5   & -0.13604  &   1  \\
19948.8210 &  &p	&  pg	  &  -47391     &  -0.14915   &  1   &     25890.5740   &  &p	&  pg	  &  -39830     & -0.21253  &   1  \\
20005.7230 &  &s	&  pg	  &  -47318.5   &  -0.22132   &  1   &     25925.5250   &  &s	&  pg	  &  -39785.5   & -0.23189  &   1  \\
20012.7920 &  &s	&  pg	  &  -47309.5   &  -0.22497   &  1   &     26029.6670   &  &p	&  pg	  &  -39653     & -0.21509  &   2  \\
20046.6120 &  &s	&  pg	  &  -47266.5   &  -0.19655   &  1   &     26127.5070   &  &s	&  pg	  &  -39528.5   & -0.21349  &   2  \\
20407.6330:&  &p	&  pg	  &  -46807     &  -0.27390   &  1   &     26352.2630   &  &s	&  pg	  &  -39242.5   & -0.21076  &   2  \\
20409.6140:&  &s	&  pg	  &  -46804.5   &  -0.25752   &  1   &     26352.2750   &  &s	&  pg	  &  -39242.5   & -0.19876  &   2  \\
20667.7760 &  &p	&  pg	  &  -46476     &  -0.24745   &  1   &     26444.6100   &  &p	&  pg	  &  -39125     & -0.20121  &   2  \\
20763.6220:&  &p &  pg	  &  -46354     &  -0.27522   &  1   &     26547.5260   &  &p	&  pg	  &  -38994     & -0.23164  &   2  \\
20769.6100 &  &s	&  pg	  &  -46346.5   &  -0.18110   &  1   &     26547.5330   &  &p	&  pg	  &  -38994     & -0.22464  &   2  \\
20789.6380 &  &p	&  pg	  &  -46321     &  -0.19229   &  1   &     26547.5410   &  &p	&  pg	  &  -38994     & -0.21664  &   2  \\
21206.5220 &  &s	&  pg	  &  -45790.5   &  -0.20203   &  1   &     26547.5490   &  &p	&  pg	  &  -38994     & -0.20864  &   2  \\
21228.5130 &  &s	&  pg	  &  -45762.5   &  -0.21485   &  1   &     26547.5560   &  &p	&  pg	  &  -38994     & -0.20164  &   2  \\
21423.7610 &  &p	&  pg	  &  -45514     &  -0.25072   &  1   &     26547.5640   &  &p	&  pg	  &  -38994     & -0.19364  &   2  \\
21481.6310:&  &s	&  pg	  &  -45440.5   &  -0.14074   &  1   &     26619.4290   &  &s	&  pg	  &  -38902.5   & -0.23397  &   2  \\
21488.6160 &  &s	&  pg	  &  -45431.5   &  -0.22840   &  1   &     26623.3410   &  &s	&  pg	  &  -38897.5   & -0.25122  &   2  \\
21542.5060 &  &p	&  pg	  &  -45363     &  -0.16916   &  1   &     26623.3620   &  &s	&  pg	  &  -38897.5   & -0.23022  &   2  \\
21544.4640 &  &s	&  pg	  &  -45360.5   &  -0.17579   &  1   &     26623.3840   &  &s	&  pg	  &  -38897.5   & -0.20822  &   2  \\
21731.7840:&  &p &  pg	  &  -45122     &  -0.28116   &  1   &     26623.4060   &  &s	&  pg	  &  -38897.5   & -0.18622  &   2  \\
21768.8060 &  &p	&  pg	  &  -45075     &  -0.19414   &  1   &     26624.5580   &  &p	&  pg	  &  -38896     & -0.21299  &   2  \\
21825.7460 &  &s	&  pg	  &  -45002.5   &  -0.22831   &  1   &     26675.2180   &  &s	&  pg	  &  -38831.5   & -0.24036  &   2  \\
21844.6320 &  &s	&  pg	  &  -44978.5   &  -0.20272   &  1   &     26711.4740   &  &s	&  pg	  &  -38785.5   & -0.13349  &   1  \\
21859.6030 &  &s	&  pg	  &  -44959.5   &  -0.16288   &  1   &     26861.5020   &  &s	&  pg	  &  -38594.5   & -0.20295  &   2  \\
21866.5970 &  &s	&  pg	  &  -44950.5   &  -0.24154   &  1   &     26913.7890    &  &p	&  pg	  &  -38528     & -0.17502  &   1  \\
22238.7070 &  &p	&  pg	  &  -44477     &  -0.23180   &  1   &     26915.3080    &  &p	&  pg	  &  -38526     & -0.22772  &   2  \\
22511.7940 &  &s	&  pg	  &  -44129.5   &  -0.22788   &  1   &     26926.7350    &  &s	&  pg	  &  -38511.5   & -0.19555  &   1  \\
22674.5270 &  &s	&  pg	  &  -43922.5   &  -0.16595   &  1   &     26929.5030    &  &p	&  pg	  &  -38508     & -0.17803  &   2  \\
22819.8350 &  &s	&  pg	  &  -43737.5   &  -0.24031   &  1   &     26979.3760    &  &s	&  pg	  &  -38444.5   & -0.20654  &   2  \\
22834.7770 &  &s	&  pg	  &  -43718.5   &  -0.22948   &  1   &     27000.2420    &  &p	&  pg	  &  -38418     & -0.16558  &   2  \\
22924.7720 &  &p	&  pg	  &  -43604     &  -0.21437   &  1   &     27002.5640    &  &p	&  pg	  &  -38415     & -0.20113  &  1    \\
23308.6560 &  &s	&  pg	  &  -43115.5   &  -0.21839   &  1   &     27026.5040    &  &s	&  pg	  &  -38384.5   & -0.22958  &  1   \\
23612.7870 &  &s	&  pg	  &  -42728.5   &  -0.21157   &  1   &     27036.3410    &  &p	&  pg	  &  -38372     & -0.21571  &  1   \\
23638.7110 &  &s	&  pg	  &  -42695.5   &  -0.22064   &  1   &     27241.8100:   &  &s	&  pg	  &  -38110.5   & -0.24664  &  1   \\

\hline
\end{tabular}
\end{table*}

\begin{table*}
\setcounter{table}{2}
\caption{Continued}
\tiny
\centering
\begin{tabular}{llccllc|llccllc}
\hline
HJD &      &&   &      &       &    &   HJD &     & &   &      &       &      \\	

(2, 400, 000+) &Error& Min. & Meth. & E & O-C(day) &Ref.&(2, 400, 000+) &Error& Min. & Meth. & E & O-C(day) &Ref. \\
  \hline

27267.8230  &  &s	&  pg	  &   -38077.5 &  -0.16671   &  1      &  34653.2960  &  &s	&  pg	  &  -28679.5  &  -0.11765  & 2  \\
27272.4990  &  &s	&  pg	  &   -38071.5 &  -0.20581   &  1      &  36788.4650  &  &s	&  pg	  &  -25962.5  &  -0.10473  & 2  \\
27313.3620  &  &s	&  pg	  &   -38019.5 &  -0.20705   &  1      &  36788.5100  &  &s	&  pg	  &  -25962.5  &  -0.05973  & 2  \\
27333.4000  &  &p	&  pg	  &   -37994   &  -0.20824   &  1      &  36790.4600  &  &p	&  pg	  &  -25960    &  -0.07436  & 2  \\
27354.5990  &  &p	&  pg	  &   -37967   &  -0.22720   &  1      &  36814.4180  &  &s	&  pg	  &  -25929.5  &  -0.08480  & 2  \\
27632.8210  &  &p	&  pg	  &   -37613   &  -0.19631   &  1      &  36840.3440  &  &s	&  pg	  &  -25896.5  &  -0.09187  & 2  \\
28027.6800  &  &s	&  pg	  &   -37110.5 &  -0.22724   &  1      &  36868.2610  &  &p	&  pg	  &  -25861    &  -0.07257  & 2  \\
28032.7880  &  &p	&  pg	  &   -37104   &  -0.22727   &  1      &  37878.4707  &0.0004  &s	&  pe	  &  -24575.5  &  -0.07381  & 3  \\
28062.7270  &  &p	&  pg	  &   -37066   &  -0.15059   &  1      &  37928.3750  &  &p	&  pe	  &  -24512    &  -0.07102  & 4  \\
28064.6470  &  &s	&  pg	  &   -37063.5 &  -0.19522   &  1      &  37935.4480  &  &p	&  pe	  &  -24503    &  -0.07068  & 4  \\
28090.6030  &  &s	&  pg	  &   -37030.5 &  -0.17229   &  1      &  37939.3762  &  0.0005&p	&  pe	  &  -24498    &  -0.07173  & 3  \\
28126.3190  &  &p	&  pg	  &   -36985   &  -0.21249   &  1      &  37941.3401  & 0.0011 &s	&  pe	  &  -24495.5  &  -0.07246  & 3  \\
28429.6470  &  &p	&  pg	  &   -36599   &  -0.22282   &  1      &  38327.5890  &  &p	&  pe	  &  -24004    &  -0.06823  & 2  \\
28668.5600  &  &p	&  pg	  &   -36295   &  -0.20841   &  1      &  38353.5219  &  &p	&  pe	  &  -23971    &  -0.06930  & 4  \\
28694.5070  &  &p	&  pg	  &   -36262   &  -0.19447   &  1      &  38662.3674  &  &p	&  pe	  &  -23578    &  -0.06309  & 4 \\
28729.5040  &  &s	&  pg	  &   -36217.5 &  -0.16783   &  1      &  38669.4397  &  &p	&  pe	  &  -23569    &  -0.06344  & 4  \\
28753.4870  &  &p	&  pg	  &   -36187   &  -0.15327   &  1      &  38672.5820  &  &p	&  pe	  &  -23565    &  -0.06454  & 2  \\
28844.5480 &  &p	&  pg	  &   -36071   &  -0.25094   &  1      &  38692.6208  &  &s	&  pe	  &  -23539.5  &  -0.06493  & 2  \\
29111.4510  &  &s	&  pg	  &   -35731.5 &  -0.14422   & 1       &  38707.5520  &  &s	&  pe	  &  -23520.5  &  -0.06489  & 2  \\
29114.5400  &  &s	&  pg	  &   -35727.5 &  -0.19862   &  2      &  38731.5204  &  &p	&  pe	  &  -23490    &  -0.06494  & 2 \\
29159.3880  &  &s	&  pg	  &   -35670.5 &  -0.14410   &  2      &  39040.3651  &  &p	&  pe	  &  -23097    &  -0.05952  & 4  \\
29161.4550  &  &p	&  pg	  &   -35668   &  -0.04173   &  2     &  39077.3002  &  &p	&  pe	  &  -23050    &  -0.05940  & 4  \\
29212.4940: &  &p	&  pg	  &   -35603   &  -0.08302   &  1      &  39338.5946  & 0.0009 &s	&  pe	  &  -22717.5  &  -0.06033  & 3  \\
29244.2330  &  &s	&  pg	  &   -35562.5 &  -0.17097   &  2      &  39340.5613  &  0.0014&p	&  pe	  &  -22715    &  -0.05825  & 3  \\
29465.4790  &  &p	&  pg	  &   -35281   &  -0.14191   &  2      &  39342.5232  &  0.0010&s	&  pe	  &  -22712.5  &  -0.06098  & 3  \\
29491.4230  &  &p	&  pg	  &   -35248   &  -0.13098   &  2      &  39355.4932  &  0.0016&p	&  pe	  &  -22696    &  -0.05751  & 3  \\
29516.5400  &  &p	&  pg	  &   -35216   &  -0.16120   &  2      &  42616.3970:  &  &s	&  vis	&  -18546.5  &  -0.04078  & 5 \\
29857.6350  &  &p	&  pg	  &   -34782   &  -0.12536   &  1      &  42660.4276  &  &s	&  pe	  &  -18490.5  &  -0.01781  & 6  \\
30181.8000  &  &s	&  pg	  &   -34369.5 &  -0.12374   &  1      &  42662.3880  &  &p	&  pe	  &  -18488    &  -0.02204  & 6  \\
30251.6300  &  &s	&  pg	  &   -34280.5 &  -0.23444   &  1      &  42671.4289  &  &s	&  pe	  &  -18476.5  &  -0.01842  & 6 \\
30259.6160 &  &s	&  pg	  &   -34270.5 &  -0.10694   &  1      &  43005.4150  &  &s	&  vis	&  -18051.5  &  -0.01883  & 7\\
30533.4650  &  &p	&  pg	  &   -33922   &  -0.12688   &  1      &  43012.5060  &  &s	&  vis	&  -18042.5  &  -0.00048  & 7 \\
30576.6840  &  &p	&  pg	  &   -33867   &  -0.12966   &  1      &  43014.4540  &  &p	&  vis	&  -18040    &  -0.01711  & 7  \\
30583.7570  &  &p	&  pg	  &   -33858   &  -0.12932   &  1      &  43016.4190  &  &s	&  vis	&  -18037.5  &  -0.01673  & 7 \\
30648.5350  &  &s	&  pg	  &   -33775.5 &  -0.18399   &  1      &  43715.4480  &  &p	&  vis	&  -17148    &  -0.00184  & 8  \\
30663.4950  &  &s	&  pg	  &   -33756.5 &  -0.15515   &  1      &  43722.5210 &  &p	&  vis	&  -17139    &  -0.00150  & 9\\
30965.6670  &  &p	&  pg	  &   -33372   &  -0.14271   &  1      &  43731.5420  &  &s	&  vis	&  -17127.5  &  -0.01778  & 9\\
30983.7290  &  &p	&  pg	  &   -33349   &  -0.15527   &  1      &  43744.5170  &  &p	&  vis	&  -17111    &  -0.00932  & 8 \\
31004.5510  &  &s	&  pg	  &   -33322.5 &  -0.15831   &  1      &  44506.7950  &  &p	&  pe	  &  -16141    &  -0.00640  & 10  \\
31247.7890  &  &p	&  pg	  &   -33013   &  -0.14107   &  1      &  44783.4139  &  &p	&  pe	  &  -15789    &  -0.00691  & 10  \\
31319.7130  &  &s	&  pg	  &   -32921.5 &  -0.12240   &  1      &  44790.4489  &  &p	&  pe	  &  -15780    &  -0.00356  & 10 \\
31323.5970  &  &s	&  pg	  &   -32916.5 &  -0.16766   &  1      &  44816.4330  &  &p	&  vis	&  -15747    &  0.00647   & 11     \\
31328.7300  &  &p	&  pg	  &   -32910   &  -0.14269   &  1      &  45216.4400  &  &p   &  pg &  -15238    &  0.01551   & 12   \\
31341.6860  &  &s	&  pg	  &   -32893.5 &  -0.15322   &  1      &  45275.3750  &  &p	&  pe	  &  -15163    &  0.01172   & 13     \\
31673.7350  &  &p	&  pg	  &   -32471   &  -0.12610   &  1      &  45568.8845  &  &s	 &  pe	&  -14789.5  &  0.00602   & 14   \\
32053.6710  &  &s	&  pg	  &   -31987.5 &  -0.14886   &  1      &  46291.4714  & 0.0010 &p	&  pe	  &  -13870    &  0.00329   & 15    \\
32381.8010  &  &p	&  pg	  &   -31570   &  -0.11149   &  1      &  46614.4600  &  &p	&  pe	  &  -13459    &  0.00729   & 16      \\
33575.5100  &  &p	&  pg	  &   -30051   &  -0.10955   &  1      &  47326.4385  &  &p	&  pe	  &  -12553    &  0.00515   & 17       \\
34636.4150  &  &p	&  pg	  &   -28701   &  -0.10286   &  2      &  47348.4413  &  &p	&  pe	  &  -12525    &  0.00413   & 17       \\
34651.3360  &  &p	&  pg	  &   -28682   &  -0.11302   &  2      &  47350.4067  &  &s	&  pe	  &   -12522.5 &    0.00491 &  17  \\

\hline
\end{tabular}
\end{table*}

\begin{table*}[!htp]
\setcounter{table}{2}
\caption{Continued}
\tiny
\centering
\begin{tabular}{llccllc|llccllc}
\hline
HJD &      &&   &      &       &    &   HJD &     & &   &      &       &      \\	
\hline
(2, 400, 000+) &Error & Min. & Meth. & E & O-C(day) &Ref.&(2, 400, 000+) &Error&  Min. & Meth. & E & O-C(day) &Ref. \\
\hline
47357.4820 & &s	&  pe	  &   -12513.5 &    0.00755     &  17        &48864.3487   & 0.0008 &p	&  pe	  &  -10596     &  0.00573     &  20 \\
47377.5175  &  &p	&  pe	  &   -12488   &    0.00386     &  17        &48865.5247   &0.0013  &s	&  pe	  &  -10594.5   &  0.00295     &  20  \\
47408.7690  &  &p	&  sp	  &   -12448   &    -0.17866    &  18        &49236.4492   & 0.0014 &s	&  pe	  &  -10122.5   &  0.00597     &  20  \\
47748.4413  &  &p	&  pe	  &   -12016   &    0.00618     &  19        &50968.8568   & 0.0020 &p	&  pe	  &  -7918      &  0.00592     &  26  \\
47761.4060  & 0.0010 &s	&  pe	  &   -11999.5 &    0.00434     &  20        &51001.8627&0.0005  &p	&  pe	  &  -7876      &  0.00609     &  26  \\
47763.3764  & 0.0008 &p	&  pe	  &   -11997   &    0.01012     &  20        &51326.4150&  &p	&  ccd	&  -7463      &  0.00210     &  27  \\
47776.3453  & 0.0009 &s	&  pe	  &   -11980.5 &    0.01248     &  20        &52130.3446   &  &p	&  pe	  &  -6440      &  0.00653     &  28 \\
47822.3117  &  &p	&  pe	  &   -11922   &    0.00662     &  21        &52144.4867   &  &p	&  pe	  &  -6422      &  0.00332     &  28  \\
47822.3191  &  &p	&  pe	  &   -11922   &    0.01402     &  21        &52172.3871   &  &s	&  pe	  &  -6386.5    &  0.00602     &  28  \\
47840.3834  &  &p	&  pe	  &   -11899   &    0.00376     &  21        &52569.6317   &  &p	&  pe	  &  -5881      &  0.00315     &  29  \\
47840.3840  &  &p	&  pe	  &   -11899   &    0.00436     &  21        &52874.1509   &  &s	&  ccd  &  -5493.5    &  0.00524     &  30  \\
48106.3953  & 0.0015 &s	&  pe	  &   -11560.5 &    0.00523     &  20    &54317.3616   &0.0020  &p	&  ccd  &  -3657      &  0.00131     &  31 \\
48106.4010  & 0.0080 &s	&  pe	  &   -11560.5 &    0.01093     &  3        &54617.5533   &0.0009  &p	&  ccd  &  -3275      &  -0.00192    &  32  \\
48115.4316  &  &p	&  pe	  &   -11549   &    0.00425     &  19        &54697.3165   & 0.0002 &s	&  ccd  &  -3173.5    &  -0.00255    &  33 \\
48132.3293  & 0.0027 &s	&  pe	  &   -11527.5 &    0.00616     &  20        &54698.4943 &0.0002  &p	&  ccd  &  -3172      &  -0.00353    &  33  \\
48145.2951  &  0.0010&p	&  pe	  &   -11511   &    0.00542     &  20        &54720.5010 &0.0004  &p	&  ccd  &  -3144      &  -0.00065    &  33 \\
48482.4214  &  &p	&  pe	  &   -11082   &    0.00182     &  3         &55083.5635   &0.0005  &p	&  pe	  &  -2682      &  -0.00112    &  34 \\
48482.4268  &0.0028  &p	&  pe	  &   -11082   &    0.00722     &  19        &55089.4572   &0.0008  &s	&  ccd  &  -2674.5    &  -0.00130    &  35 \\
48482.4272  &  &p	&  pe	  &   -11082   &    0.00762     &  20        &55447.4114   &0.0010  &p	&  ccd  &  -2219      &  -0.00205    &  36  \\
48484.3885  & 0.0054 &s	&  pe	  &   -11079.5 &    0.00429     &  3         &55774.3275   &0.0007  &p	&  ccd  &  -1803      &  0.00020     &  37  \\
48484.3966  &0.0036  &s	&  pe	  &   -11079.5 &    0.01239     &  20        &55777.4684   &0.0008  &p	&  ccd  &  -1799      &  -0.00230    &  37 \\
48488.3169  & 0.0014 &s	&  pe	  &   -11074.5 &    0.00344     &  20        &55779.4327   &0.0002  &s	&  ccd  &  -1796.5    &  -0.00263    &  37 \\
48489.4935  & 0.0018 &p	&  pe   &   -11073   &    0.00126     &  20        &55793.5788   &0.0003  &s	&  ccd  &  -1778.5    &  -0.00184    &  37  \\
48500.5030  &  &p	&  pe	  &   -11059   &    0.00885     &  22        &55799.4728   &  &p	&  ccd	&  -1771      &  -0.00172    &  22  \\
48504.4299  & 0.0005 &p	&  pe	  &   -11054   &    0.00650     &  3         &55873.3412   &  &p	&  ccd	&  -1677      &  -0.00328    &  22 \\
48532.3265  & 0.0007 &s	&  pe	  &   -11018.5 &    0.00540     &  3         &57191.2159   & 0.0003 &p	&  ccd	&  0          &  0.00000     &  38 \\
48559.4382  &  &p  &  pe   &   -10984   &    0.00526     &  23        &57214.3977   &0.0064  &s & ccd   &  29.5       &  -0.00082    &  39 \\
48559.4387  &  &p	&  pe	  &   -10984   &    0.00576     &  23        &57294.1623   &0.0005  &p	&  ccd	&  131        &  0.00002     &  38\\
48598.3360  &  &s	&  pe	  &   -10934.5 &    0.00345     &  24        &57324.0210   &0.0006  &p	&  ccd	&  169        &  -0.00123    &  38\\
48823.4853   &  &p	&  pe	  &  -10648     &  0.00656      &  25        &57644.64990  & 0.0002 &p	&  ccd	&  577        &  -0.00183    &  40 \\

\hline
\end{tabular}
\tablecomments{0.86\textwidth}{
(1)\citet{strbau1968}. (2)   \citet{kondo1966}.  (3)  \citet{wolf1992}. (4)   \citet{kruseman1968}. (5)   \citet{bonneville1975}. (6) \citet{dumdin1976}. (7)   \citet{bramun1979}. (8)   \citet{bramun1981}. (9) GEOS EB 13. (10)   \citet{aslanherczeg1984}. (11) BAV-M 34. (12)BAV-M 36.   (13)\citet{diethelm1983}.  (14)
 \citet{scarfe1984}  (15) \citet{pohl1987}. (16)BBSAG 80. (17) \citet{kespoh1989}. (18)\citet{ahn1992}. (19) \citet{wunder1992}. (20) \citet{muyesseroglu1996}. (21)BAV-M 56     (22)B.R.N.O. data. (23)BAV-M 60. (24) BBSAG 99.  (25)\citet{blattler1992b} . (26)   \citet{nelson1998}. (27)Rotse. (28) \citet{derman2003}. (29)\citet{nelson2003} (30)VSOLJ 42.
 (31)  \citet{brat2007}. (32)\citet{brat2008}. (33)\citet{yilmaz2009}. (34)  \citet{hubscher2010}.   (35)\citet{gokay2010}. (36)\citet{gokay2012}. (37) \citet{liania2011}.     (38) This study. (39)  \citet{hubscher2016}. (40)\citet{jurysek2017}.
 }
\end{table*}

\begin{table*}[!ht]
\normalsize
\caption{Orbital parameters of V1073 Cyg for two O-C diagram analysis}
\label{table4}
\centering
\begin{tabular}{lcc}
\hline
                       Parameter                            & Value                                 \\
\hline
  \hline
Revised epoch, $ \triangle T_0$(day)                          &0.026834($\pm$0.004848)                    \\
Revised period, $ \triangle P_0(days)$                          &3.0261848 ($\pm0.4545760)\times 10^{-6}$    \\
Rate of the linear decrease, $\beta$(\period)     & -1.04$(\pm0.18)\times10^{-10}$  &           \\
Eccentricity, $e_3$                                            &0.26($\pm$0.15)                              \\
Longitude of the periastron passage,$\omega_3(^{\circ})$       &197.9  ($\pm$25.3  )                          \\
Time of periastron passage, $T_3$                               &2453082.8 ($\pm$ 2111.1)                      \\
Semi-amplitude, A (day)                                         &0.028($\pm$0.002)                             \\
Orbital period, $P_3$ (years)                                    &82.7($\pm$3.6)                                \\
Orbital semi-major axis, ${a_{12}}^{'} \sin{i^{'}}$(AU)         & 4.84($\pm$0.34)                              \\
 Mass function, $f(m)(M_\odot)$                                 &0.017($\pm0.004$)                              \\
 Mass, $M_3(i^{'} =90^{\circ})(M_\odot)$                         &0.511                                         \\
orbital radius,$a_3(i^{'} =90^{\circ})(AU)$                      &22.368                                        \\
  \hline
 \end{tabular}
\end{table*}

\section{DISCUSSION AND CONCLUSION}
V1073 Cyg is a short-period W UMa type eclipsing binary with a period of $P=0.7858506 day$. Our new multi-color light curves analysis found that V1073 Cyg is a shallow contact binary, with a fill-out factor of $f(\emph{fill-out})= 0.124(\pm0.011)$ , an orbit inclination angle of $i=68.4(\pm2.9)^\circ$ and $M_1=1.810 (\pm0.004) M_\odot$, $M_2=0.549(\pm0.001) M_\odot$.

The O-C analysis demonstrated that the orbital period of V1073 Cyg is undergoing a continuous decrease and a cyclic variation simultaneously. The period decrease rate $\dot P=-1.04(\pm0.18)\times 10^{-10}$\period is approximately consistent with $\dot P=-7.8\times{10}^{-10}$ \period of \citet{sezer1993} and $\dot P=-8.8\times{10}^{-10}$ \period\ of \citet{yang2000}. More high-precision data (20 CCD data and more pe data) with a longest time span (119 yeas) were used in our analysis, the results derived from our O-C analysis will be more reliable.
The long-term period changes could be explained by the combined effects of angular momentum loss (AML) and mass transfer between two components. With period decrease, V1073 Cyg will evolve from shallow contact stage to deep contact stage, then it will merge into a rapid rotational single star, like V1309 Sco \citep{zhu2016}. Long-term period variation (no matter decrease or increase) are very common for W UMa-type contact binaries, like V502 Oph\citep{derman1992}, V417 Aql \citep{qian2003}, V1191 Cyg\citep{zhu2011}, EP And  \citep{liao2013}, TY UMa\citep{samec2000}, DD Ind \citep{samec2016} and so on.

The cyclic variation curve fits the O-C diagram very well in Figure \ref{fig5}, and this is the first time to detect a cyclic variation for V1073 Cyg. The period of the cyclic variation is $P_3=82.7(\pm3.6) years$ and the corresponding amplitude $A=0.028(\pm0.002) day$. There are two possible ways to interpret the cyclic variation, i.e. the magnetic activity of one or both components \citep{applegate1992}, and the LTTE through the presence of a tertiary companion. In consideration of the short period and fast rotational velocity of about $160 km\cdot s^{-1}$, and the constitution of a F0V type primary component and a late type secondary component, it is probable that magnetic activity cycles \citep{applegate1992,lanza1998,lanza1999} in the two components may appear in V1073 Cyg.
Using the equation given by \citep{rovithis2000}:
\begin{eqnarray}
\centering
\Delta P &=& \sqrt{2[1-\cos(2\pi P/P_3)]}\times A
\end{eqnarray}
In the formula, $P_3$ is the period of the cyclic oscillation, we can obtain the value of  $\displaystyle\frac{\Delta P}{P}$. The variation of the quadruple momentum $\Delta Q$ causing such cyclic variation can be calculated with following equation \citep{lanza2002}:
\begin{eqnarray}
\centering
\displaystyle\frac{\Delta P}{P}&=& -9\displaystyle\frac{\Delta Q}{Ma^2}
\end{eqnarray}
in which $a$ is the separation between the two components. As $M_1=1.810 M_\odot$,
$M_2=0.549 M_\odot$ and $a=5.172 R_\odot$, combining the above two equations, $\Delta Q_1=3.6\times 10^{52} g {cm}^2$ and $\Delta Q_2= 1.7\times 10^{52} g {cm}^2$ can be obtained for the components, and the values of $\Delta Q_1$ and $\Delta Q_2$ are typical ones for close binaries, for which the $\Delta Q$ should be about $10^{51}\sim 10^{52} g {cm}^2$ \citep{lanza1999}, so the Applegate mechanism can be used to explain the cyclic period variation.

According to \citet{liao2010}, the LTTE due to the existence of a distant third companion may be the plausible reason of the periodic variation of binaries. Third body plays an important role in the evolution of contact binaries. It will cause the periodic variation of period, and also will take away the angular momentum of system leading to the orbital contraction process and then promote the evolution of contact binaries \citep{Tokovinin2006,qian2013,qian2014,qian2017}. In Table \ref{table2}, $M_1=1.810(\pm0.122) M_\odot$, and $M_2=0.549(\pm0.037) M_\odot$, then the mass function of the tertiary companion can be calculated with the equation:
\begin{eqnarray}
 \centering
 f(m)&=&\displaystyle\frac{4\pi^2}{G{P_3}^2}\times ({a_{12}}^{'}\sin{i^{'}})^3=\displaystyle\frac{(M_3\sin{i^{'}})^3}{(M_1+M_2+M_3)^2}
\end{eqnarray}
In the formula, $G$ is the gravitational constant, and $P_3$ is the orbital period of tertiary component. ${a_{12}}^{'} \sin{i^{'}}=4.84(\pm0.34)(AU)$ leads to $f(m)=0.017(\pm0.004)M_\odot$, and if $i^{'}=90^{\circ}$, then $M_3=0.511M_\odot$ and $a_3=21.368 AU$ can be calculated out. This is the first time to detect the sign of the existence of a third body for V1073 Cyg. Our O-C analysis show that the mass of the third body is just little less than the mass of the second component, but we failed to fit the light curve with third light. No third body has been detected from the light curve analysis in the previous researches, and the radial velocity curves \citep{fitzgerald1964,pribulla2006} didn't show any evidence of potential third body. One possible explanation is that the third body may exist as a white dwarf. In fact, the mass of the third body is around the mass peak distribution of dwarfs.

In our opinion, the LTTE will be more plausible than the the magnetic activity mechanism to interpret the cyclic variation. Because, comparing with the magnetic cycles shown in solar-type single stars and close binaries \citep{maceroni1990,bianchini1990}, the period $82.7 years$ of cyclic variation of V1073 Cyg is much long. LTTE may cause a strict cyclic variation of period not a simple period oscillation, which can be verify by more new CCD times of light minima data.

As we have mentioned that V1073Cyg is a special doubtful Am binary because of the fast rotation velocity. From our light curve and O-C curve analysis, we know that V1073Cyg is a shallow contact binary and the period is undergoing a long-term decrease. The mass transfer and loss occurring in this shallow contact binary will lead to the orbit shrinking. Because of the synchronous rotation, the components of V1073 Cyg will rotate faster, which makes  V1073 Cyg more interesting.

\begin{acknowledgements}
This work is partly supported by the Key Science Foundation of Yunnan Province (No. 2017FA001), Chinese Natural Science Foundation (Nos.11573063, 11325315 and U1631108) ,  CAS "Light of West China" Program, the research fund of Sichuan University of Science and Engineering (grant number 2015RC42).

New CCD photometric observations of the  system were obtained with the 85 $cm$ telescope at Xinglong station of National Astronomical Observatory and the 60 $cm$ telescope in Yunnan Observatories.
\end{acknowledgements}





\newpage
\begin{thebibliography}{}
\bibitem[Abt(1961)]{abt1961}Abt, H. A., 1961, ApJS, 6, 37
\bibitem[Abt(1965)]{abt1965}Abt, H. A., 1965, ApJS, 11, 429
\bibitem[Abt et al.(1969)]{abt1969} Abt, H. A., Bidelman, W.P., 1969, \apj,158, 1091
\bibitem[Abt \& Moyd(1973)]{abt1973}Abt, H. A.,  Moyd, K. I. 1973, \apj,182, 809
\bibitem[Ahn et al.(1992)]{ahn1992}Ahn, Y. S.; Hill, G.; Khalesseh, B., 1992, A\&A, 265, 597A
\bibitem[Applegate (1992)]{applegate1992}Applegate J. H., 1992, ApJ, 385, 621
\bibitem[Aslan \& Herczeg(1984)]{aslanherczeg1984}Aslan, Z., Herczeg, T., 1984, IBVS, 2478, 1A
\bibitem[Beoche \& Grebel(2016)]{beoche2016}Boeche, C., Grebel, E. K., 2016, A\&A, 587A, 2B
\bibitem[Blattler(1992b)]{blattler1992b}Blattler, E .1992b , BBSAG, 101, 3
\bibitem[Bianchini(1990)]{bianchini1990}Bianchini A., 1990, AJ, 99, 1941
\bibitem[Bonneville et al.(1975)]{bonneville1975}Bonneville, T., Chetanneau, A., Desprez, F. et al. 1975, BBSAG, 23, 1B
\bibitem[Br\'{a}t et al.(2007)]{brat2007} Br\'{a}t, L., Zejda, M., Svoboda, P. 2007, OEJV, 74, 1
\bibitem[Br\'{a}t et al.(2008)]{brat2008} Br\'{a}t, L., \v{S}melcer, L., Ku\`{e}\'{a}kov\'{a}, H. et al. 2008, OEJV, 94,1B
\bibitem[Braune \& Mundry(1979)]{bramun1979}Braune, W., J., Mundry, E. 1979, AN, 300, 165
\bibitem[Braune \& Mundry(1981)]{bramun1981}Braune, W., J., Mundry, E. 1981, AN, 302, 53
\bibitem[Conti(1970)]{conti1970}Conti, P. S. 1970,  \pasp, 82, 781
\bibitem[Cox(1999)]{cox1999} Cox, Arthur N., 1999, Allen's Astrophysical Quantities, Springer-Verlag
\bibitem[Derman \& Demircan(1992)]{derman1992} Derman, E., Demircan, O., 1992, AJ, 103,1658D
\bibitem[Derman \& Kalci(2003)]{derman2003}Derman, E., Kalci, R. 2003, IBVS, 5439, 1

\bibitem[Diethelm et al.(1983)]{diethelm1983}Diethelm, R.,  Elias, D. P., Germann, R., et al. 1983, BBSAG, 64, 1
\bibitem[Dumitrescu \& Dinescu(1976)]{dumdin1976}Dumitrescu, A., Dinescu, R. 1976, IBVS, 1116, 1
\bibitem[Ekmek\c{c}i(2012)]{ekmekci2012}Ekmek\c{c}i, F., Elmasli, A., Yilmaz, M., 2012, NewA, 17, 603E
\bibitem[Fitzgerald(1964)]{fitzgerald1964}Fitzgerald, M.P., 1964, PDDO.2, 417
\bibitem[Gokay et al.(2012)]{gokay2012}Gokay, G., Demircan, Y., Gursoytrak, H., et al. 2012, IBVS, 6039, 1
\bibitem[Hubrig et al.(2010)]{hubrig2010} Hubrig S., Gonz\'aez J. F., Sch\'oller M., et al. 2010, ASPC, 435, 257
\bibitem[Hubscher et al.(2010)]{hubscher2010}Hubscher, J., Lehmann, P. B., Monninger, G.,  et al. 2010, IBVS, 5941, 1
\bibitem[Hubscher(2016)]{hubscher2016}Hubscher, Joachim, 2016, IBVS, 6157, 1H
\bibitem[Irwin(1952a)]{irwin1952a}Irwin, J. B. 1952a, \apj,116, 211
\bibitem[Jafari et al.(2006)]{jafari2006}Jafari, M., Khalesseh, B., Pazhouhesh, R., 2006, Ap\&SS, 306, 29J
\bibitem[Jury\u{s}ek et al.(2017)]{jurysek2017} Jury\u{s}ek, J., Ho\u{n}kov\u{a}, K., \u{S}melcer, L.,  et al., 2017, OEJV, 179, 1J
\bibitem[Keskin \& Pohl(1989)]{kespoh1989}Keskin, V., Pohl, E. 1989, IBVS, 3355, 1
\bibitem[Kondo(1966)]{kondo1966}Kondo, Y., 1966, AJ, 71, 54
\bibitem[Kruseman(1968)]{kruseman1968} Kruseman, P. 1968, BANS, 2, 377
\bibitem[Gokay et al.(2010)]{gokay2010} Gokay, G., Demircan, Y., Terzioglu, Z. et al. 2010, IBVS, 5922, 1G
\bibitem[Lanza et al.(1998)]{lanza1998}Lanza, A. F., Rodon\`{o}, M., \& Rosnor, R. 1998, MNRAS, 296, 893
\bibitem[Lanza \& Rodon\`{o}(1999)]{lanza1999}Lanza A. F., Rodon\`{o} M., 1999, A\&A, 349, 887
\bibitem[Lanza \& Rodon\`{o}(2002)]{lanza2002}Lanza A. F., Rodon\`{o} M., 2002, Astron. Nachr., 323, 424
\bibitem[Leung \& Schneider(1978)]{leung1978}Leung, K.C.,  Schneider, D.P. 1978, \apj,222, 917
\bibitem[Liao \& Qian(2010)]{liao2010}Liao, W.P.,  Qian, S.-B. 2010, MNRAS, 405, 1930
\bibitem[Liao et al.(2013)]{liao2013}Liao, W-P.,  Qian, S.B.,  Li, K., et al. 2013, AJ, 146, 79L
\bibitem[Liakos \& Niarchos(2011)]{liania2011}Liakos, A., Niarchos, P. 2011, IBVS, 6005, 1
\bibitem[Maceroni et al.(1990)]{maceroni1990}Maceroni C., Bianchini A., Rodono M., et al. 1990, A\&A, 237, 395
\bibitem[Morgan et al.(1943)]{morgan1943} Morgan, W. W.; Keenan, P. C.; Kellman, E. 1943, An atlas of stellar spectra(Chicago: University of Chicago press)
\bibitem[Morris \& Naftilan(2000)]{morris2000}Morris, S. L., Naftilan, S. A. 2000,  \pasp, 112, 852
\bibitem[Muyesseroglu et al.(1996)]{muyesseroglu1996}Muyesseroglu, Z., Gurol, B., Selam, S. O. 1996, IBVS, 4380, 1
\bibitem[Nelson(1998)]{nelson1998}Nelson, R. H. 1998, IBVS, 4621, 1
\bibitem[Nelson(2003)]{nelson2003}Nelson, R. H. 2003, IBVS, 5371, 1
\bibitem[Pohl et al.(1987)]{pohl1987}Pohl, E.,  Akan, M. C., Ibanoglu, C., et al. 1987, IBVS, 3078, 1
\bibitem[Pribulla et al.(2006)]{pribulla2006}Pribulla, T., Rucinski, S. M., Lu, W., et al. 2006, AJ, 132, 769
\bibitem[Qian(2001b)]{qian2001b}Qian, S. B. 2001b, MNRAS, 328, 635
\bibitem[Qian(2003)]{qian2003} Qian, S. B. 2003, A\&A, 400, 649
\bibitem[Qian et al.(2013)]{qian2013}Qian, S. B., Liu, N. P., Li, K.,  et al. 2013, ApJS, 209, 13Q
\bibitem[Qian et al.(2014)]{qian2014} Qian, S. B., Zhou, X.,  Zola, S., et al. 2014, AJ, 148, 79Q
\bibitem[Qian(2015)]{qian2015}Qian, S. B., Han, Z. T., Fern\'{a}ndez Laj\'{u}s, E. et al. 2015, ApJS, 221, 17Q
\bibitem[Qian et al.(2017)]{qian2017}Qian, S. B., He, J. J., Zhang, J., et al. 2017, RAA, 17, 87
\bibitem[Renson \& Manfroid(2009)]{renson2009}Renson, P., Manfroid, J. 2009, A \& A, 498, 961
\bibitem[Roman et al.(1948)]{roman1948}Roman, N. G., Morgan, W. W., \& Eggen, O. J. 1948, \apj,107, 107
\bibitem[Rovithis-Livaniou et al.(2000)]{rovithis2000}Rovithis-Livaniou H., Kranidiotis A. N., Rovithis P., et al. 2000, A\&A, 354, 904
\bibitem[Samec et al.(2000)]{samec2000}Samec, R. G., Stoddard, M. L., Faulkner, D. R. 2000, AAS Meeting Abstracts, 196, 4601S
\bibitem[Samec et al.(2016)]{samec2016}Samec, R. G., Norris, C. L., Van Hamme, W., et al. 2016, AJ, 152, 219S
\bibitem[Scarfe et al.(1984)]{scarfe1984}Scarfe, C. D., Forbes, D. W., Delaney, P. A., Gagne, J. 1984, IBVS, 2545, 1
\bibitem[Sezer(1993)]{sezer1993}Sezer, C. 1993, Ap\&SS, 208, 15
\bibitem[Sezer(1994)]{sezer1994}Sezer, C. 1994, Ap\&SS, 215, 153
\bibitem[Sezer(1996)]{sezer1996}Sezer, C. 1996, Ap\&SS, 245, 89
\bibitem[Smalley et al.(2014)]{smalley2014}Smalley, B., Southworth, J., Pintado, O. I. et al., 2014, A\&A, 564A, 69S
\bibitem[Strohmeier(1960)]{strohmeier1960} Strohmeier, W. 1960, Veroeffentlichungen der Remeis-Sternwarte zu Bamberg, Nr. 27. Ver\"{a}nderlichen-Colloquium Bamberg 1959. Bamberg, 1960, p.1
\bibitem[Strohmeier(1962)]{strohmeier1962}Strohmeier, W. 1962, IBVS, 9
\bibitem[Strohmeier \& Bauernfeind(1968)]{strbau1968}Strohmeier, W., Bauernfeind, H. 1968, BamVe, 7, 72
\bibitem[Titus \&\ Morgan(1940)]{titus1940}Titus, J., \& Morgan, W. W. 1940, \apj,92, 256
\bibitem[Tokovinin et al.(2006)]{Tokovinin2006}Tokovinin, A.; Thomas, S.; Sterzik, M.; Udry, S. 2006, A\&A, 450, 681T
\bibitem[Van Hamme \& Wilson(2007)]{van2007}Van H. W., Wilson, R. E. 2007, AJ, 661, 1129
\bibitem[Wilson(1979)]{wilson1979}Wilson, R. E., 1979, AJ, 234, 1054
\bibitem[Wilson(1990)]{wilson1990}Wilson, R. E., 1990, AJ, 356, 613
\bibitem[Wilson(2008)]{wilson2008}Wilson, R. E., 2008, AJ, 672, 575
\bibitem[Wilson(2012)]{wilson2012}Wilson, R. E., 2012, AJ, 144, 73
\bibitem[Wilson \& Devinney(1971)]{wildev1971}Wilson, R.E., Devinney, E.J. 1971, AJ, 166, 605
\bibitem[Wilson et al.(2010)]{wilson2010}Wilson, R.E., Van Hamme, W., Terrell, D. 2010, AJ, 723, 1469
\bibitem[Wolf \& Diethelm(1992)]{wolf1992}Wolf, M., Diethelm, R. 1992, AcA, 42, 363
\bibitem[Wunder et al.(1992)]{wunder1992}Wunder, E., Wieck, M., Kilinc, B. et al. 1992, IBVS, 3760, 1W
\bibitem[Yang \& Liu(2000)]{yang2000}Yang, Y., Liu, Q. 2000, Ap\&SS, 274, 799
\bibitem[Yilmaz et al.(2009)]{yilmaz2009}Yilmaz, M. et al. 2009, IBVS, 5887,1
\bibitem[Zhu et al.(2011)]{zhu2011}Zhu, L. Y.; Qian, S. B.; Soonthornthum, B.; He, J. J.; Liu, L., 2011, AJ, 142, 124Z
\bibitem[Zhu et al.(2016)]{zhu2016}Zhu, L. Y.; Zhao, E. G.; Zhou, X., 2016, RAA, 16, 68Z



\end{thebibliography}
\end{document}